%% file: main.tex
\documentclass{article}

\usepackage{arxiv}

\usepackage[utf8]{inputenc} % allow utf-8 input
\usepackage[T1]{fontenc}    % use 8-bit T1 fonts
\usepackage{hyperref}       % hyperlinks
\usepackage{url}            % simple URL typesetting
\usepackage{booktabs}       % professional-quality tables
\usepackage{amsfonts}       % blackboard math symbols
\usepackage{nicefrac}       % compact symbols for 1/2, etc.
\usepackage{microtype}      % microtypography
\usepackage{lipsum}
\usepackage{fancyhdr}       % header
\usepackage{graphicx}       % graphics
\graphicspath{{media/}}     % organize your images and other figures under media/ folder
\usepackage{enumitem}
\usepackage{cite}

\usepackage{amsmath,amssymb,amsfonts}
\usepackage{algorithmic}
\usepackage{graphicx}
\usepackage{textcomp}
\usepackage{xcolor}
\usepackage{subcaption}
\usepackage{balance}
\usepackage[nopostdot,nogroupskip,nonumberlist]{glossaries}
\usepackage{physics}
 
\let\ket=\undefined

\usepackage{tikz}
\usetikzlibrary{shapes.arrows}
\usetikzlibrary{shapes.arrows}
\usetikzlibrary{positioning}

\usepackage[braket, qm]{qcircuit}
\newcommand{\killsectionspace}{\vspace{-0.1cm}}
\newcommand{\killsubsectionspace}{\vspace{-0.05cm}}
  
\title{A Case for Noisy Shallow Gate-based Circuits in Quantum Machine Learning
\thanks{
\textbf{P. Selig, N. Murphy, A. Sundareswaran R, D. Redmond, S. Caton. ``A Case for Noisy Shallow Gate-based Circuits in Quantum Machine Learning''. International Conference on Rebooting Computing (ICRC). IEEE, 2021.}} 
}

\author{
  Patrick Selig \\
  School of Computer Science\\
  Centre for Quantum Engineering,\\ Science, and Technology\\
  University College Dublin \\
  Dublin\\
  \texttt{patrick.kappen@ucdconnect.ie} \\
   \And
  Niall Murphy\\
  Equal1 Labs \\
  Dublin\\
  \texttt{niall.murphy@equal1.com} \\
  \And
  Ashwin Sundareswaran R\\
  Equal1 Labs \\
  Dublin\\
  \texttt{ashwin@equal1.com} 
  \And
  David Redmond\\
  Equal1 Labs \\
  Dublin\\
  \texttt{david.redmond@equal1.com} 
  \And
  Simon Caton \\
  School of Computer Science\\
  Centre for Quantum Engineering,\\ Science, and Technology\\
  University College Dublin \\
  Dublin\\
  \texttt{simon.caton@ucd.ie} \\
}

\newacronym{vqml}{VQML}{Variational Quantum Machine Learning}
\newcommand{\vqml}{\gls{vqml} {}}
\newacronym{pcs}{PCs}{Principal Components}
\newcommand{\pcs}{\gls{pcs} {}}

\begin{document}
\maketitle

\begin{abstract}
There is increasing interest in the development of gate-based quantum circuits for the training of machine learning models. Yet, little is understood concerning the parameters of circuit design, and the effects of noise and other measurement errors on the performance of quantum machine learning models. In this paper, we explore the practical implications of key circuit design parameters (number of qubits, depth etc.) using several standard machine learning datasets and IBM's Qiskit simulator. In total we evaluate over 6500 unique circuits with $\mathbf{n \approx 120700}$ individual runs. We find that in general shallow (low depth) wide (more qubits) circuit topologies tend to outperform deeper ones in settings without noise. We also explore the implications and effects of different notions of noise and discuss circuit topologies that are more / less robust to noise for classification machine learning tasks. Based on the findings we define guidelines for circuit topologies that show near-term promise for the realisation of quantum machine learning algorithms using gate-based NISQ quantum computer. 
\end{abstract}

% keywords can be removed

\newif\iffinal
%\finaltrue

\iffinal
  \newcommand\patrick[1]{}
  \newcommand\simon[1]{}
  \newcommand\david[1]{}
  \newcommand\niall[1]{}
  \newcommand\ashwin[1]{}  
\else
  \newcommand\patrick[1]{{\color{red}[***Patrick: #1]}}
  \newcommand\simon[1]{{\color{blue}[***Simon: #1]}}
  \newcommand\david[1]{{\color{orange}[***David: #1]}}
  \newcommand\niall[1]{{\color{magenta}[***Niall: #1]}}
  \newcommand\ashwin[1]{{\color{green}[***Ashwin: #1]}}
\fi   

\input{tex_files/Introduction}
\input{tex_files/Related_Work}
\input{tex_files/Approach}
\input{tex_files/Results}
\input{tex_files/Conclusion}

\section{Acknowledgements}
\killsectionspace
This publication has emanated from research conducted with the financial
support of Science Foundation Ireland under Grant number 18/CRT/6183. For the purpose
of Open Access, the author has applied a CC BY public copyright licence to any
Author Accepted Manuscript version arising from this submission. This work has been carried out using the ResearchIT Sonic cluster which was funded by UCD IT Services and the Research Office.

%Bibliography
\bibliographystyle{unsrt}  
\bibliography{references}

\end{document}

%% file: tex_files/Introduction.tex
\section{Introduction}
\label{section:introduction}
\killsectionspace
Quantum Computing has shown a lot of promise for many domains as it exhibits properties like entanglement and superposition that do not exist in classical scenarios. One area that is expected to profit greatly from this is machine learning. Current and near term Noisy Intermediate-Scale Quantum (NISQ) era quantum computers are, however, still very error prone but the class of quantum machine learning models based on variational principles is expected to perform well despite the presence of noise \cite{varcirc, mlqc_noise, noise_resilience_vqa}. In \vqml models, model parameters are classical while the computation of the prediction based on the input, i.e. training process, is performed on the quantum computer. 

Concrete implementations of \vqml algorithms can be quite diverse (see for example \cite{9259971, hardwareansatz, 9259966, quantum_chemistry}). Typically, a quantum circuit is the representation of quantum algorithms based on gates operating on quantum bits (qubits). We  focus on the view of quantum computations as circuits, but this does not restrict our results as they could be viewed as operators acting on a quantum state as well. However, there are (as of yet) no clear rules or guidelines on how to best design circuits specifically for machine learning tasks. There are approaches to design variational quantum circuits using reinforcement learning or simulated annealing (e.g. \cite{Fosel2021QuantumCO}) but these techniques are sample inefficient and therefore would not be applicable when access to real devices is scarce and or expensive as the simulation of quantum circuits is not feasible for higher number of qubits. An engineer intending to design a \vqml circuit would therefore benefit greatly from having guidelines helping in their design. In a similar vain, deep circuits have a higher capacity in principal over shallower circuits given a fixed number of qubits, it is not clear whether the higher capacity can be taken advantage of. To help address these challenges, this paper seeks to establish a set of core design principles for \vqml; specifically for classification tasks. We investigate the effects of key architecture properties on machine learning model performance: noise apparent in the system (inducing a range of errors into the model), and the number of qubits used. We also investigate circuit-specific properties, i.e., the length of the critical path through a quantum circuit: its depth.%., and comment briefly on the choice of classical optimisation strategy to steer the training process. 

To derive an observational dataset of different \vqml circuits, we use IBM's Qiskit \cite{aleksandrowicz2019qiskit} simulator and six ``simple'' datasets from the machine learning literature for classification tasks. To form design guidelines, a large design space of candidate circuits is explored under different conditions (noise), topology properties (number of qubits, circuit depth), and methods of handling the dataset (how many features are used and preprocessed).
We use simulation models for this study to allow precise control of noise properties, which will carry forward to hardware design requirements. The guidelines outlined in this work hold true in simulations even without noise as the majority of our experiments show.

Our main findings are that shallow variational quantum circuits outperform deeper ones when performing classification, even when disregarding the effects of noise. Increasing the number of qubits in the circuits in general also improves model performance. A small number of features benefits model performance.
Shallow circuits are also more resilient to different kinds of noise. As such, our results currently indicate that shallow circuits with larger numbers of qubits are well suited to machine learning tasks in the presence of noise.

The paper is structured as follows: In \autoref{section:related_work} we discuss relevant related work. In \autoref{section:approach} we provide an overview of our experimental approach. Sections \ref{section:Noise_free_results} and \ref{sec:noise_results} discuss the main results without and with noise (i.e. error) respectively. Finally, section \ref{sec:limitations} presents high-level implications (limitations) as areas of further work and \ref{section:conclusion} presents our main conclusions and itemizes a set of design guidelines for \vqml circuits.

%% file: tex_files/Related_Work.tex
\section{Related Work}
\label{section:related_work}
\killsectionspace
\subsection{Quantum Machine Learning}
\killsubsectionspace
There are different Quantum Machine Learning approaches that seek to exploit quantum speed-up. For example an adapted form of Grover’s search algorithm \cite{qsvm} and the HHL matrix inversion algorithm \cite{qhnn}. These works are dependent on the quantum computations being error free, which is not possible in NISQ era of QC. An alternative direction makes use of variational circuits \cite{variational_survey} which are suitable for the NISQ era. Quantum Neural Networks \cite{tdqnn, cqnnnnp, 9259966} are similar to the approach taken in this paper as they apply successive layers of classically parameterized quantum operations. In one of the two approaches taken in \cite{qefc} a \vqml circuit is trained to learn a feature map of the data, as is done in \cite{fhs}. In contrast, our work directly learns a quantum classifier. Variational quantum circuits are used as classifiers in \cite{pqcmlm, ccqc, edfevqc} but these works make the explicit distinction between an encoder part in the circuit and a classifier part. In contrast our work does not make this distinction and is most similar to the approach taken in \cite{9259971}. Instead of  finding a circuit for a given problem, we explore a circuit design space to derive circuit design guidelines.

A problem that strongly influences the performance in variational quantum circuits for machine learning are barren plateaus \cite{barrenplat}. It was shown in \cite{shallowbarren} that shallow circuits can  suffer from barren plateaus when a global cost function is used. There have been different approaches aiming to reduce the influence of barren plateaus, for example layer wise training \cite{layerwise},  spatially or temporally correlating gate layers \cite{correlation} or through low depth measurements \cite{lowdepth}. \cite{transition} have shown that there is an abrupt transition in the trainability of variational quantum circuits and that the layer wise pre-training is not always possible. As we explore the impact of the depth of a quantum circuit and number of qubits, this paper can be viewed as an experimental study of barren plateaus in training \vqml circuits on a variety of datasets.

\subsection{Variational Forms}
\killsubsectionspace
As this work explores the impact of the circuit architecture on the performance of quantum machine learning models, it is related to the analysis of variational forms \cite{varcirc, vqs} and the representational power of parametric quantum circuits \cite{expres}.  \cite{expres} proposes a technique to find a maximally expressive ansatz for a minimum number of parameters. In contrast our work explores an ansatz based on fixed blocks from which circuits are sampled and evaluated. % in various experimental settings. 
The variational forms used in \cite{vqs, vqste, vqsgp} are motivated and specialized for the physical problem at hand. In \cite{Kyriienko2020SolvingND}, a framework for solving nonlinear differential equations with various variational quantum circuits was proposed. The work used two variational forms preceded by a feature encoding called a feature map as is done in the machine learning community. The variational form we use is based on the hardware efficient ansatz in \cite{hardwareansatz}. Similar to our work, \cite{VQML_MNIST} uses PCA to reduce the number of features in the MNIST\cite{MNIST} dataset to 8.

\subsection{Optimization of Quantum Machine Learning Models}
\killsubsectionspace
In this work we made use of the COBYLA optimizer which has shown promising results when the influence of noise is small \cite{classoptim} and it has proven successful in similar experiments in \cite{9259971, Terashi2021} while still being fast to compute \cite{classoptim}. In the presence of noise in the computations other optimizers like ImFil and NOMAD should be preferred \cite{classoptim}. Although these optimizers were not originally designed for the use with quantum computers, recent works have proposed optimizers that are designed for the optimization of variational quantum circuits. \cite{icanss} proposes iCANS, an optimizer that performs stochastic gradient descent by using the parameter shift rule and estimating each partial derivate seperately. This was expanded in \cite{quantum_chemistry} who proposed Rosalin an optimizer that combines iCANS\cite{icanss} with two different sampling schemes. In contrast our approach with COBYLA does not compute gradients. While COBYLA builds a surrogate model of the loss with trust regions, \cite{modeloptim} proposes two model based optimizers that have shown promise, especially in the presence of noise, using stochastic gradients instead of trust regions.

% \subsection{Frameworks for Quantum Computing}
% \begin{itemize}
%     \item Qiskit
%     \item Q\#
%     \item Cirq
%     \item TensorFlow Quantum
%     \item PennyLane   
% \end{itemize}

%\subsection{Summary}

%% file: tex_files/Approach.tex
\section{Approach}
\label{section:approach}
\killsectionspace
We define a design space of basic blocks that act on qubits: % that allows the exploration of a wide range of circuits of varying depth and number of qubits. 
a $\frac{\pi}{2}$ rotation around the $X$-axis followed by a parametric rotation around the $Z$-axis followed by another $\frac{\pi}{2}$ rotation around the $X$-axis, as suggested by \cite{9259971}; see \autoref{fig:basic_block}. The parametric $Z$ rotation can either be used as input for a feature or as a trainable parameter. Blocks are repeated multiple times depending on the number of features in the dataset and degree of feature repetition in the circuit (a design parameter). After each block a Controlled $Z$ ($CZ$) gate can be applied between different qubits. The depth of the circuit, the number of features and number of qubits are dependent variables when the all features are to be used. An example 3 qubit circuit receiving 5 features is illustrated in \autoref{fig:complete_circuit}, where $\omega_i$ represents the $i$-th feature, $\theta_j$ represents the $j$-th trainable parameter, and the connections between qubits are $CZ$-gates. A measurement (not shown) is applied at the end of the circuit on each qubit.

\newcommand{\Rxpi}{\mathrm{R}_X(\frac{\pi}{2})}
\newcommand{\Rz}[1]{\mathrm{R}_Z(#1)} 

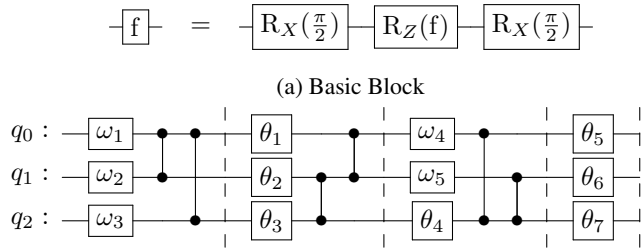
\begin{figure}[h]
\centering
\begin{subfigure}[b]{\linewidth}
%\includegraphics[width=.2\textwidth]{figures/Base_Block.jpg}
% qc = qiskit.QuantumCircuit(1)
% qc.rx(np.pi / 2, 0)
% qc.rz(5,0)
% qc.rx(np.pi / 2, 0)
% print(qc.draw(output="latex_source")) %then copy out the equation part and change "5" to "f1"
\newcommand{\f}{\mathrm{f}}
\[
\Qcircuit @C=.5em @R=0em @!R {
& \gate{\f} & \qw & \push{\rule{.3em}{0em}=\rule{.3em}{0em}} & & \gate{\Rxpi} & \qw & \gate{\Rz{\f}} & \qw & \gate{\Rxpi} & \qw
}
\]

% quantikz seems much nicer but its not compatible with the qiskit one which seems more useful. 
% \begin{quantikz}
% &\gate{R_x(-\pi/2)} & \gate{R_z(f1)} & \gate{R_x(\pi/2)} &\\
% \end{quantikz}
\caption{Basic Block}
\label{fig:basic_block}
\end{subfigure}
\\
\begin{subfigure}[b]{\linewidth}
% \includegraphics[width=.49\textwidth]{figures/qc_example.png}
% \begin{equation*}
%     \Qcircuit @C=1.0em @R=0.2em @!R {
% 	 	\lstick{ {q}_{0} :  } & \gate{\mathrm{H}} & \ctrl{1} \barrier[0em]{1} & \qw \barrier[0em]{1} & \qw & \meter & \qw & \qw & \qw\\
% 	 	\lstick{ {q}_{1} :  } & \qw & \targ & \qw & \qw & \qw & \meter & \qw & \qw\\
% 	 	\lstick{c:} & \lstick{/_{_{2}}} \cw & \cw & \cw & \cw & \cw & \cw & \cw & \cw\\
% 	 	\lstick{meas:} & \lstick{/_{_{2}}} \cw & \cw & \cw & \cw & \dstick{_{_{0}}} \cw \cwx[-3] & \dstick{_{_{1}}} \cw \cwx[-2] & \cw & \cw\\
% 	 }
% \end{equation*}
    \begin{equation*}
        \Qcircuit @C=1.0em @R=0.2em @!R {
        \lstick{ {q}_{0} :  } & \gate{\mathrm{\omega_{1}}} & \ctrl{1}  & \ctrl{1} \barrier[0em]{2} & \qw & \gate{\mathrm{\theta_{1}}} & \qw       & \ctrl{1} \barrier[0em]{2} & \qw & \gate{\mathrm{\omega_{4}}} & \ctrl{1}  &\qw \barrier[0em]{2} & \qw & \gate{\mathrm{\theta_{5}}} \barrier[0em]{2} & \qw \\
        \lstick{ {q}_{1} :  } & \gate{\mathrm{\omega_{2}}} & \ctrl{-1} & \qw                       & \qw & \gate{\mathrm{\theta_{2}}} & \ctrl{1}  & \ctrl{-1}                 & \qw & \gate{\mathrm{\omega_{5}}} & \qw       &\ctrl{1}             & \qw & \gate{\mathrm{\theta_{6}}}                  & \qw \\
        \lstick{ {q}_{2} :  } & \gate{\mathrm{\omega_{3}}} & \qw       & \ctrl{-1}                 & \qw & \gate{\mathrm{\theta_{3}}} & \ctrl{-1} & \qw                       & \qw & \gate{\mathrm{\theta_{4}}} & \ctrl{-1} & \ctrl{-1}           & \qw & \gate{\mathrm{\theta_{7}}}                  & \qw\\
        % 	 	\lstick{c:} & \lstick{/_{_{3}}} \cw & \cw & \cw & \cw & \cw & \cw & \cw & \cw &\cw& \cw &\cw & \cw &\cw &\cw\\
        }
    \end{equation*}
\caption{Example 3-qubit circuit operating on 5 features ($\omega_i$). There are 7 trainable parameters ($\theta_j$). Black dots indicate $CZ$-gates between two qubits. Dashed lines demarcate layers.}
\label{fig:complete_circuit}
\end{subfigure}
\caption{Illustrative examples of our circuit components.}
\vspace{-.3cm}
\end{figure}

To explore a large range of candidate circuits the following workflow was established. Given a dataset and (classical) optimizer, we: 1) create a number of candidate circuits by randomly sampling the available design space with the parameters: number of qubits, the entanglement pattern, the number of blocks and the configuration of blocks, i.e., whether they are used as feature input or a trainable parameter; 2) compile these into an executable Qiskit circuit; 3) derive a number of noise models to be applied to circuits. The noise models are defined through the two decoherence times and the gate duration; 4) compile the noise models into Qiskit executable noise model; and 5) train all permutations of circuit designs and noise models and evaluate their performance. From a machine learning perspective, this entails a holdout strategy (i.e. using a training set and unseen test set for evaluation) repeated 10 times to estimate the variance in the results. Steps 3 and 4 are only performed when noise models are defined for the experiments (in Section \ref{sec:noise_results}), otherwise they will be executed without any noise as is the case in Section \ref{section:Noise_free_results}.

\subsection{Number of qubits}
\killsubsectionspace
The number of qubits in a quantum circuit is, disregarding hardware constraints, a design parameter when building a \vqml circuit. The number of qubits has a big influence on how the features can be inserted into the algorithm, since feature rotations can be applied to each qubit it allows the maximal depth to be lower. This also has influence on the feature interactions since it depends on the $CZ$-gates patterns between the qubits. Only if qubits are connected directly or indirectly by $CZ$-gates can the feature values interact. To evaluate the influence of the number of qubits on the performance of a machine learning system, we sampled circuits of different depths while varying between 2-10 qubits. We restrict our circuit design space such that the number of qubits is always less than or equal to the total number of features in the dataset.

\subsection{Circuit depth}
\label{ssec:circuit_depth}
\killsubsectionspace
In a quantum circuit the depth defines the critical (i.e. longest) path. There are various design decisions that influence the overall depth of the quantum circuit when designing it for \vqml. One can choose to repeat features or use more or less entanglement. These decisions influence the performance of the final classifier. The minimal and maximal depth for each dataset varies as it depends on the number of qubits used and the number of features in the dataset. The minimal depth for a circuit is bound by $\lceil\frac{numFeatures}{numQubits}\rceil$. 
For example, a dataset with 4 features over 4 qubits requires 6 rotations to add two basic blocks (see~\autoref{fig:basic_block}), one for the feature and one for the parameter, adding, for example, one $CZ$-gate and the measurement gate results in a minimal depth of 8. Due to those constraints, the range of circuit depth varies between 8 and 142 on multiple data sets.

\subsection{Circuit Design Space}
\killsubsectionspace
Since the overall design space of \vqml circuits is very big, we draw random samples from a defined set of possible circuit architectures. This defined set comprises circuits that apply a parameterized block (see \autoref{fig:complete_circuit}) on each qubit followed by between 1 and $n$ (a predefined constant) $CZ$-gates%. This is repeated a specific number of times depending on 
. This set of circuits was chosen as they can represent a wide range of different functions, even though they cannot compute all functions, while at the same time being minimal in depth~\cite{9259971}. The design space chosen falls in the space of hardware efficient ans\"atze \cite{hardwareansatz} where Euler rotations are replaced with the basic blocks, introduced previously, and the entanglers were selected at random.

\subsubsection{Controlled Z Gate}
\killsubsectionspace
Overall using $CZ$-gates is not a restriction as other controlled gates would be possible by changing the basis and measuring along another dimension.

\subsubsection{Controlled Z Gate patterns}
\killsubsectionspace
Between each parameterized $\theta$ block of rotation gates, we insert a layer of $CZ$-gates configured as follows. First, we select the number of $CZ$-gates uniformly from $\{1,2,3\}$.
The pairs of qubits to which they are applied are sampled at random. There are never two $CZ$-gates applied to the same set of qubits in one layer. In our experiments we maximally applied either 3 $CZ$-gates or connect each qubit with each other qubit via a $CZ$-gate.

\subsubsection{Layer structure}
\killsubsectionspace
We refer to a layer as the combination of a set of parameterized blocks and the corresponding $CZ$-gates. Each layer adds a set of features or trainable parameters, where the number of features or parameters is equal to the number of qubits. If not otherwise specified, all features of an example in a dataset will be used at least once in the model. New parameters are used for padding when the number of features is not a multiple of the number of qubits in the circuit. Feature layers and parameter layers are alternated. The first layer and last layer can either be a feature or parameter layer. Therefore, the total number of layers in the circuit depends on the number of features and number of qubits. Features can be repeated to essentially double the total number of layers in the model which has proven to be beneficial in some works \cite{9259971}. An example circuit can be found in \autoref{fig:complete_circuit}.

\subsubsection{Output mapping}
\killsubsectionspace
To map the output distribution of bit strings to classes the bit string space is split up into equal ranges, i.e., the number of classes. Left over bit strings are ignored. Each range is mapped to a class. This way the minimal and maximal number of qubits is independent of the number of classes as long as the number of bit string combinations is equal to or larger than the number of classes. The output distribution is normalized to add up to one and the probabilities of each class is computed by summing the probabilities of each associated bit string. The classifier loss is the negative log likelihood of the softmax of the output:
\begin{equation}
   \mathcal{L}_\theta(x,y)=-log\frac{e^{P_y}}{\sum_{k\in K}e^{P_k}}
\end{equation}
Here $K$ represents the set of all classes, $x$ represents the input feature, and $y$ the label of the input. $P_y$ represents the probability of target class $y$ while $P_k$ represents the probability of class $k$. In each optimization step the mean of the whole training set is computed and optimized over.

\subsection{Errors in Quantum Computing}
\killsubsectionspace
The computations in NISQ quantum computers are error prone, it is therefore important to evaluate the influence of errors in the computations on the performance of a classifier. This is especially important since the error created through noise affects not only the final performance of the classifier but also the computation of the loss function as it depends on the erroneous computations from the circuit. The error in a quantum computer is often not normally distributed, but is biased towards a relaxation from an excited state into a base state and towards a relaxation of the superposition. Therefore our noise model uses the two thermal relaxation times T1 and T2 representing the time for the relaxation of an excited state to a base state (T1), and the relaxation of a superposition, also called dephasing time (T2) respectively.

\begin{figure}[h]
\centering
\includegraphics{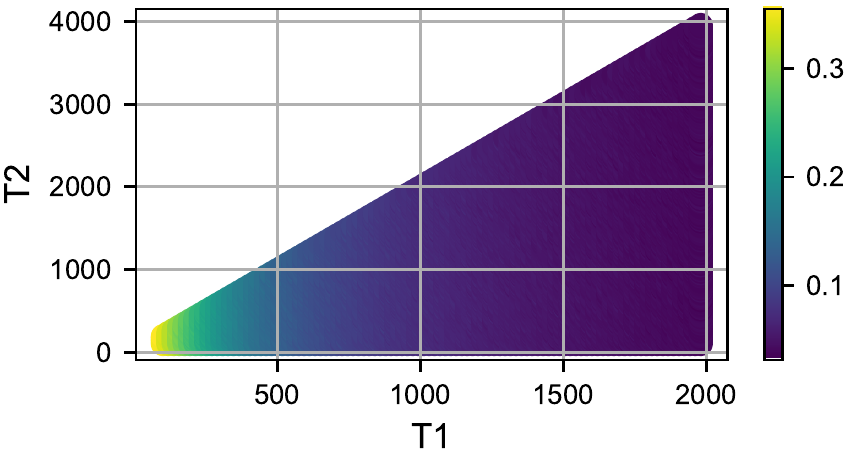}
\caption{T1 vs. T2 relaxation times in ns. The Color represents the error value. T1 and T2 must satisfy $T2 \leq 2T1$.}
\label{noise_example}
\vspace{-.25cm}
\end{figure}

Here, we focus mainly on the noise in the $CZ$-gate and the measurement gate. In order to do this we created noise models for the simulations. The noise modeled the two thermal relaxation times T1 and T2. T1 was in the range of 100ns to 80000ns and T2 was set in the range of 100ns to 140000ns and the gate time was 300ns for the $CZ$-gate. For the measurement operations the T1 time was in the range of 100ns to 80000ns and T2 was set in the range of 100ns to 140000ns and the gate time was 1000ns. As the thermal relaxation times do not represent errors themselves we estimated the overall error by taking 10000 shots on a noisy circuit containing only the respective operation and compute the mean distance to the correct value. The error for the $CZ$-gate ranges from 0 to 35\% and for the measurement operation from 0\% to 50\%. How the T1 and T2 relaxation times map to the error can be seen in \autoref{noise_example} for the $CZ$-gate.

\subsection{Datasets}
\killsubsectionspace
% \begin{itemize}
%     \item Iris\footnote{\url{https://archive.ics.uci.edu/ml/datasets/iris}}
%     \item Wine\footnote{\url{https://archive.ics.uci.edu/ml/datasets/wine}}
%     \item Ecoli\footnote{\url{https://archive.ics.uci.edu/ml/datasets/ecoli}}
%     \item Vertebral Column\footnote{\url{http://archive.ics.uci.edu/ml/datasets/vertebral+column}}
%     \item Glass Identification\footnote{\url{https://archive.ics.uci.edu/ml/datasets/Glass+Identification}}
%     \item Breast Cancer Wisconsin \footnote{\url{https://archive.ics.uci.edu/ml/datasets/Breast+Cancer+Wisconsin+(Diagnostic)}}
% \end{itemize}

\begin{figure*}[htp]
\centering
  \begin{subfigure}[t]{150pt}
    \includegraphics[width=1\textwidth]{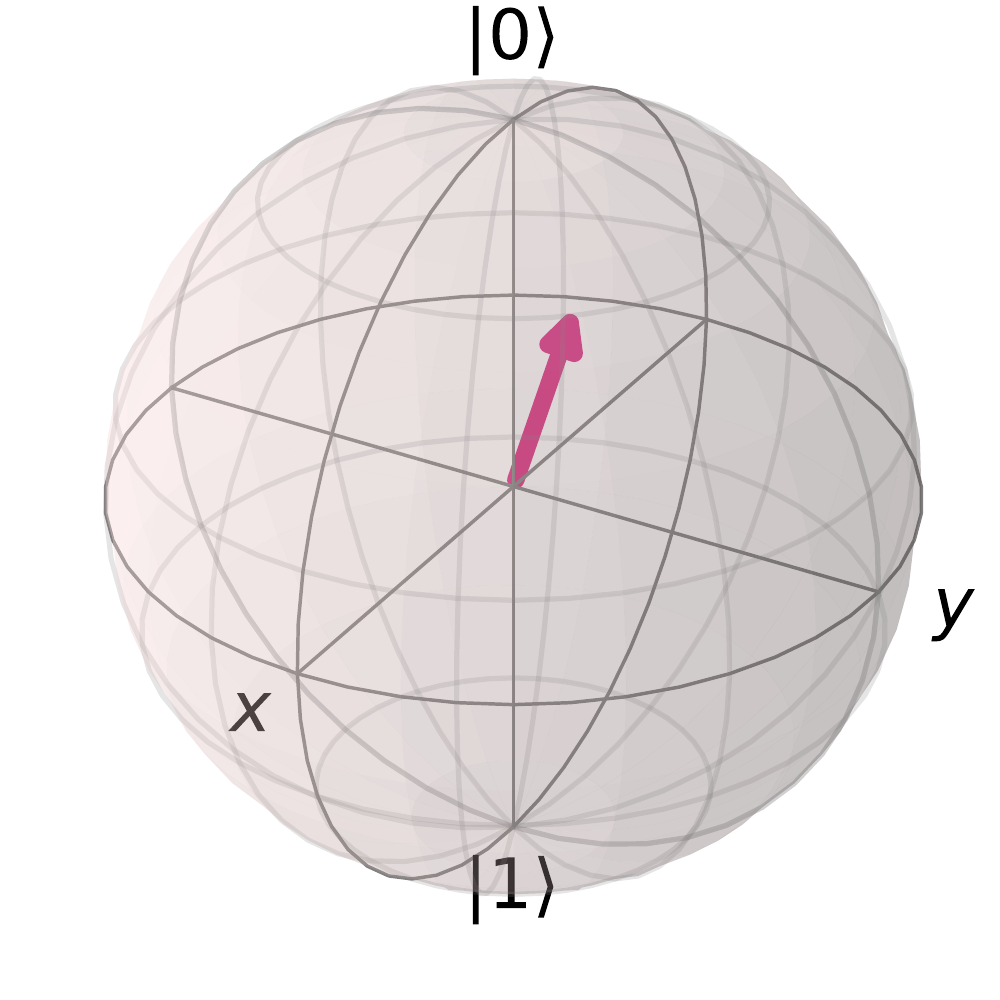}
\caption{Minimal feature value of -5.}
\label{min_bloch}
  \end{subfigure}
  \begin{subfigure}[t]{150pt}
\includegraphics[width=1\textwidth]{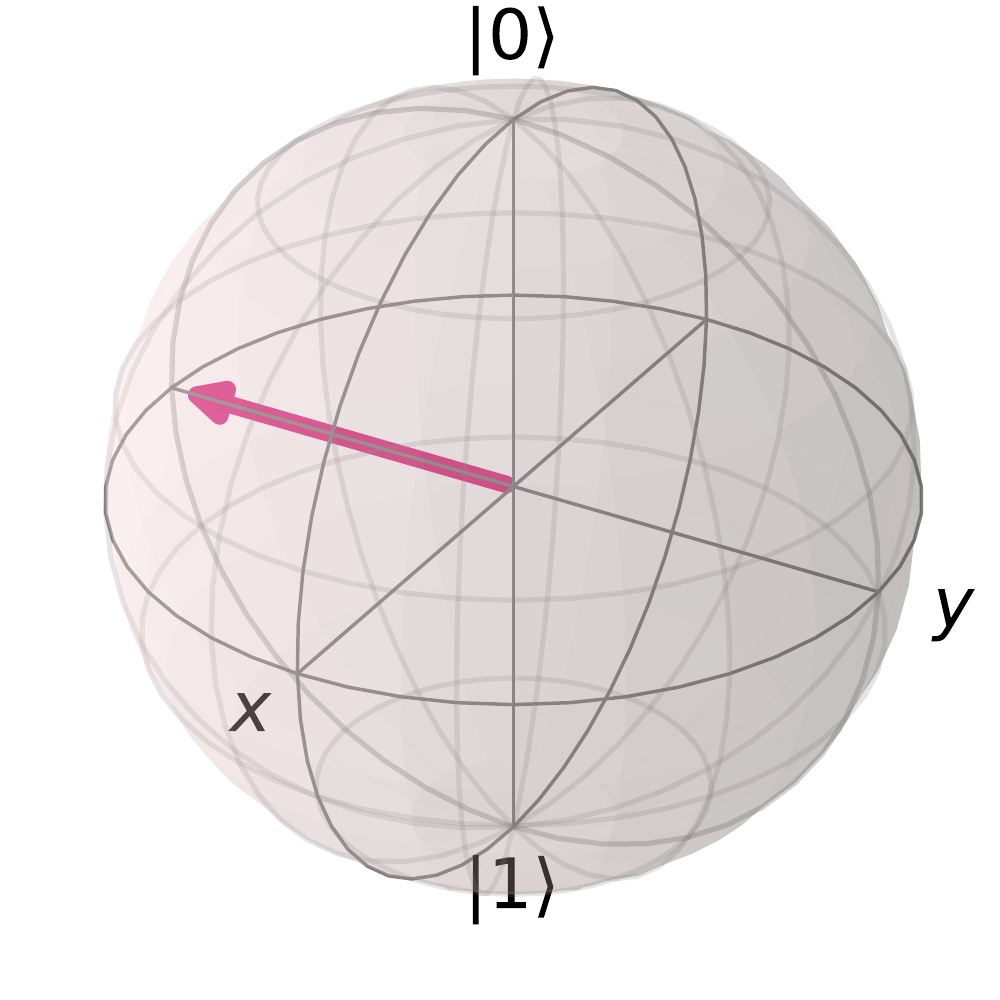}
\caption{Feature value of 0.}
\label{middle_bloch}
  \end{subfigure}
  \begin{subfigure}[t]{150pt}
\includegraphics[width=1\textwidth]{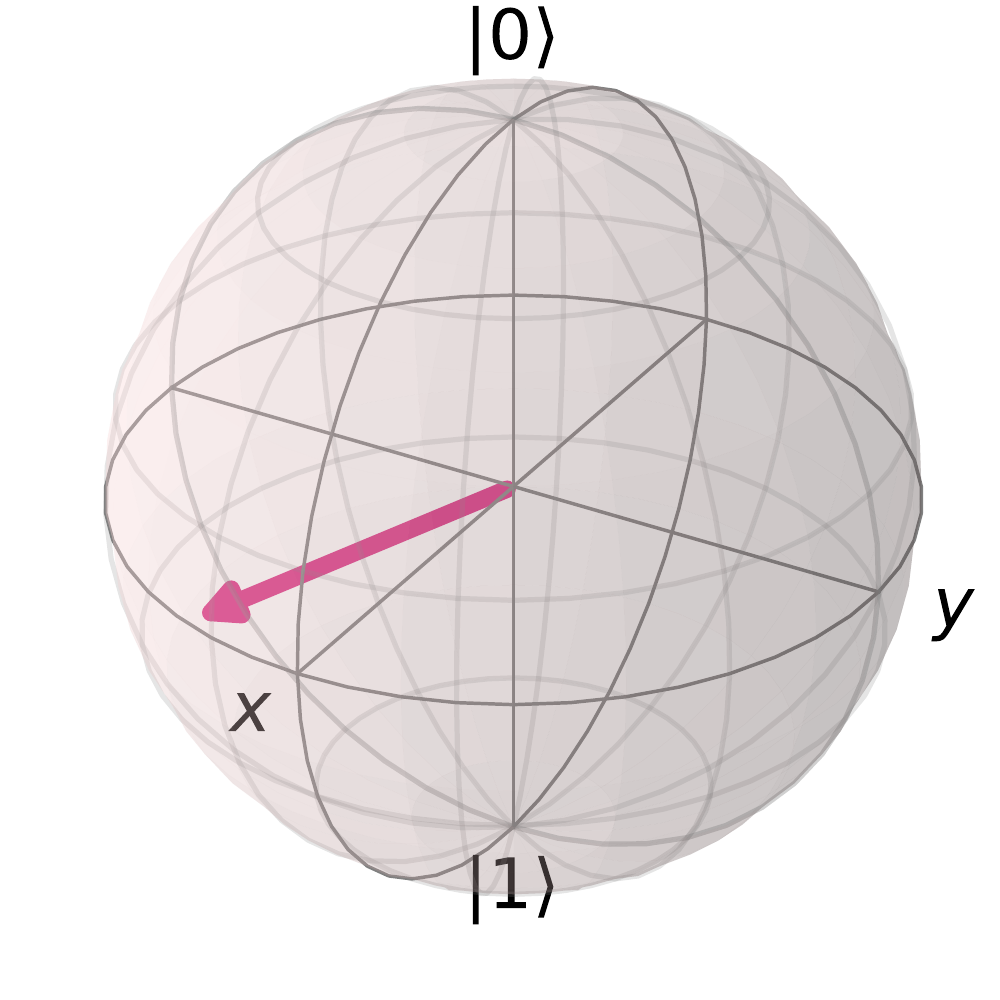}
\caption{Maximal feature value of 5}
\label{max_bloch}
  \end{subfigure}
  \hfill %%
  \caption{\label{fig:bloch}Visualization of the feature values -5, 0, 5 mapped to the respective $\Rz{\omega}$ rotation with the parameters $\alpha=0.05$ and $q=3$. The $\Rz{\omega}$ is applied after an initial $\Rxpi$ (see \autoref{fig:basic_block}), which means all arrows (mapped feature values) lie in the equatorial plane (z=0). Initially all qubits are in the state $\ket{0}$.}
  \vspace{-.3cm}
\end{figure*}

To derive guidelines for \vqml circuit design, a range of real valued datasets within a multivariate classification setting were selected. This restriction allows us to evaluate the influence of the number of qubits and the depth of a circuit on a specific class of machine learning problems. This is simple enough to evaluate a wide range of circuits (datasets have only a few hundred examples), yet complex enough to still pose a challenging problem. By allowing only real valued features, we remove the challenges of exploring different encodings for inputting integers and categorical features. %Thus, we can train a large set of circuits as each dataset contains only a few hundred examples.
The real valued features are applied by the basic blocks in \autoref{fig:basic_block} through parameterized rotations around the $Z$-axis.
The real valued features are initially centred (mean of zero) and scaled (standard deviation of one), then rescaled as suggested by \cite{9259971}:
\begin{equation}
    f(x)=(1-\frac{\alpha}{2})\frac{\pi}{q}W
\end{equation}
Here $W$ represents the normalized features, $q$ the quantile of a Gaussian approximation and $\frac{\alpha\pi}{q}$ the angular gap between the extreme points of the distribution. For all experiments the angular gap $\alpha\pi$ is set to $\frac{\pi}{10}$ and the quantile $q$ to 3. A visualization of encoding the feature values -5, 0 and 5 with the aforementioned parameters can be found in \autoref{fig:bloch}. Principal component analysis (PCA) was used to reduce the number of features for the Wine (8 and 12 components), Breast Cancer (12 and 16 components) and Glass (8 components) datasets. All other dataset-specific information is shown in \autoref{DatasetTable}.

\begin{table}[ht]

\begin{center}
\begin{tabular}{|c|c|c|c|c|}
\hline
%\textbf{Dataset}&\multicolumn{4}{|c|}{\textbf{Properties}} \\\cline{2-5} 
 & \textit{\bf No. of} & \textit{\bf No. of} & \textit{\bf Train} & \textit{\bf Test} \\ 
\textbf{Dataset} & \textbf{\textit{Features}}& \textbf{\textit{Classes}}& \textbf{\textit{Size}} & \textbf{\textit{Size}}\\
\hline
Iris\cite{iris} & 4 & 3 & 90 & 60\\
Wine\cite{wine} & 13 & 3 & 108 & 70\\
Ecoli\cite{ecoli} & 7 & 8 & 266 & 70\\
Vertebral Column 1\cite{vertebral} & 6 & 2 & 240 & 70\\
Vertebral Column 2\cite{vertebral} & 6 & 3 & 240 & 70\\
Glass Identification\cite{glass} & 9 & 7 & 144 & 70\\
Breast Cancer Wisconsin\cite{cancer} & 32 & 2 & 449 & 120\\
\hline

\end{tabular}
\caption{Datasets used in this analysis.}
\label{DatasetTable}
\end{center}
\vspace{-.5cm}
\end{table}

%to move to refs to open up a bit more space
% \footnotetext[1]{\url{https://archive.ics.uci.edu/ml/datasets/iris}}
% \footnotetext[2]{\url{https://archive.ics.uci.edu/ml/datasets/wine}}
% \footnotetext[3]{\url{https://archive.ics.uci.edu/ml/datasets/ecoli}}
% \footnotetext[4]{\url{https://archive.ics.uci.edu/ml/datasets/iris}}
% \footnotetext[5]{\url{https://archive.ics.uci.edu/ml/datasets/Glass+Identification}}
% \footnotetext[6]{\url{https://archive.ics.uci.edu/ml/datasets/Breast+Cancer+Wisconsin+(Diagnostic)}}

\subsection{Training Setup}
\killsubsectionspace
All \vqml models (circuits) were trained using the COBYLA optimizer. In our training regime, we used a conservative initial change to the variables of 1.0 and an absolute constraint violations tolerance of 0.0002. These values have proven to be quite good in \cite{9259971} and allow the values to vary more at the beginning while at the same time requiring a high final accuracy. The circuits were optimized for a maximum of 200 epochs. The number of shots were increased as the training progressed. This makes sure that the influence of sampling is reduced as the model gets better at predicting the results while at the same time minimizes the computation costs. A fixed schedule was used: 250 samples for epochs 1-20, 500 samples for 21-50, and 750 samples for 51-200. When testing the circuit on the unseen test set, 300 samples were taken for each example. Despite using simulation, we sampled the circuit instead of using a state vector representation. This makes the results more realistic, as a state vector is not available on real hardware, only measurements of the state along a specific dimension; in our case, along the z eigenvector of the state. The parameters of the circuit were initialized by sampling from a uniform distribution in the range of $[-\pi, \pi]$. 

%% file: tex_files/Results.tex
\section{Noise-Free Circuit Performance}
\killsectionspace
\label{section:Noise_free_results}
In this section we present the results of training various circuit topologies without any noise, i.e. they represent an idealised scenario of error free computations. Although this is unrealistic with modern NISQ hardware, it acts as a baseline for discussion. However, as this section will illustrate, even without noise we find that shallow circuits with more qubits tend to outperform deeper circuits with fewer qubits. Overall training variational quantum classifiers is not easy, as even shallow circuits with a high(er) number of qubits exhibit a high variance in the final circuit performance after training.

\subsection{Number of Qubits vs. Circuit Depth}
\begin{figure*}[htbp]
\centering
  \begin{subfigure}[t]{150pt}
    \includegraphics[width=1\textwidth]{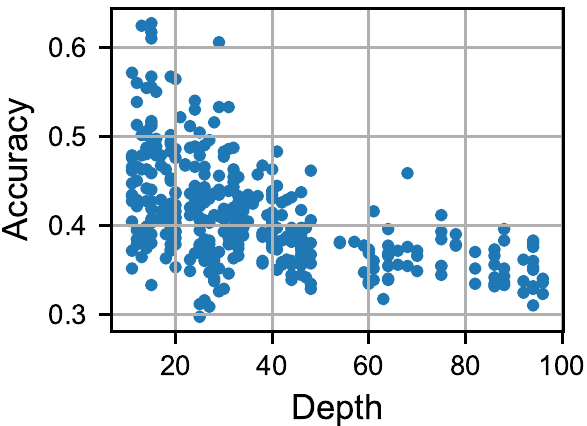}
\caption{Accuracy vs. depth for $p=12$~\pcs.}
\label{wine_pca_12_depth_acc}
  \end{subfigure}
  \begin{subfigure}[t]{150pt}
\includegraphics[width=1\textwidth]{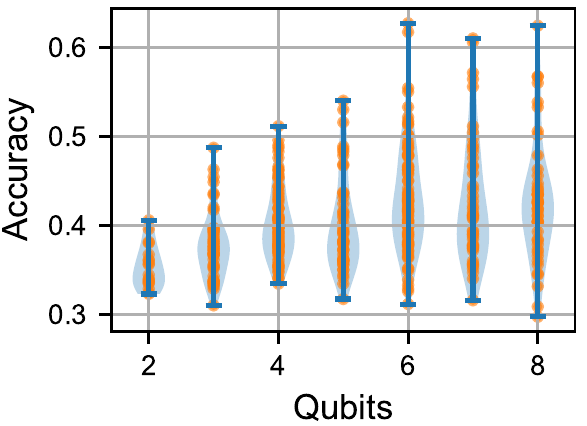}
\caption{Accuracy vs. No. qubits for $p=12$~\pcs.}
\label{wine_pca_12_qubits}
  \end{subfigure}
  \begin{subfigure}[t]{150pt}
\includegraphics[width=1\textwidth]{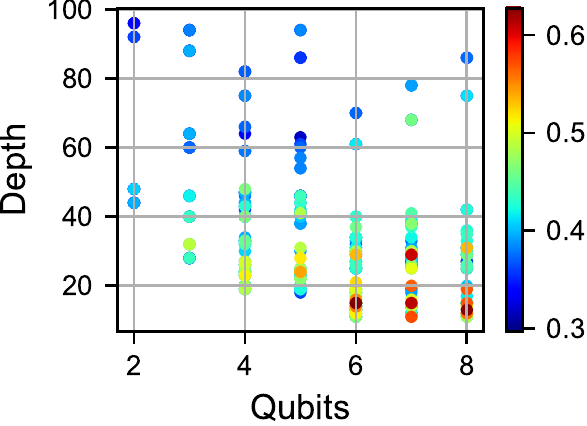}
\caption{Depth vs. No. qubits colour-coded by Accuracy for $p=12$~\pcs.}
\label{wine_pca_12_depth_bits}
  \end{subfigure}
  \hfill %%
  \caption{\label{fig:wine}Relationship between accuracy, number of qubits, and depth on the first $12$ principal components (PCs) of the Wine dataset. Each point represents the mean of 10 training runs without noise.}
  \vspace{-.3cm}
\end{figure*}

\begin{figure*}[htbp]
\centering
  \begin{subfigure}[t]{150pt}
    \includegraphics[width=1\textwidth]{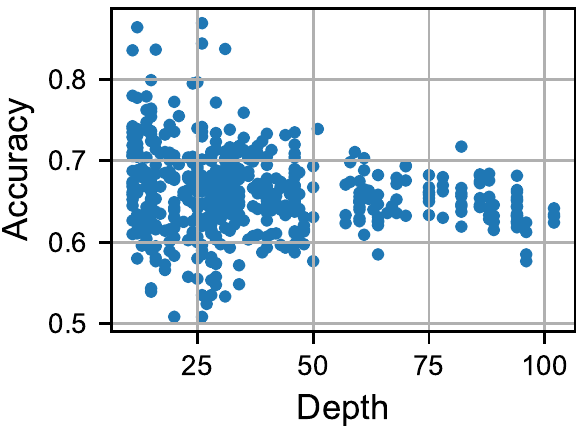}
\caption{Accuracy vs. depth for $p=12$~\pcs.}
\label{breast_cancer_pca_12_depth_acc}
  \end{subfigure}
  \begin{subfigure}[t]{150pt}
\includegraphics[width=1\textwidth]{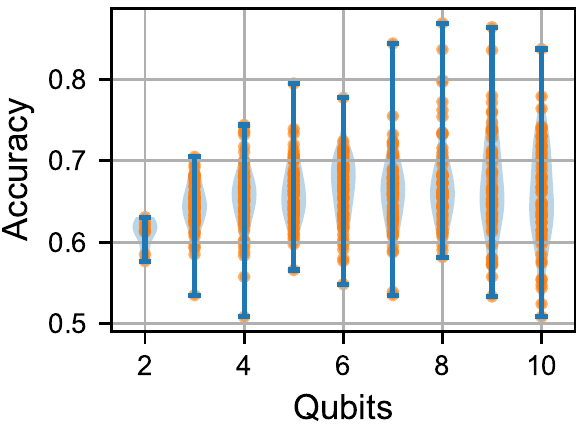}
\caption{Accuracy vs. No. qubits for $p=12$~\pcs.}
\label{breast_cancer_pca_12_qubits}
  \end{subfigure}
  \begin{subfigure}[t]{150pt}
\includegraphics[width=1\textwidth]{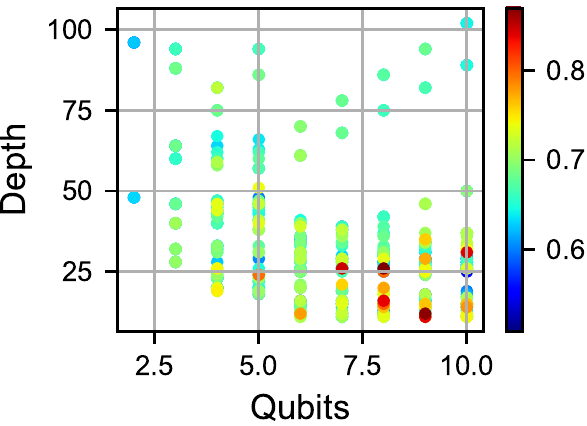}
\caption{Depth vs. No. qubits colour-coded by Accuracy for $p=12$~\pcs.}
\label{breast_cancer_pca_12_depth_bits}
  \end{subfigure}
  \hfill %%
  \caption{\label{fig:breast_cancer}Relationship between accuracy, number of qubits, and depth on the first $12$ principal components (PCs) of the Breast Cancer Wisconsin dataset. Each point represents the mean of 10 training runs without noise.}
  \vspace{-.3cm}
\end{figure*}

\begin{figure*}[htbp]
\centering
  \begin{subfigure}[t]{150pt}
    \includegraphics[width=1\textwidth]{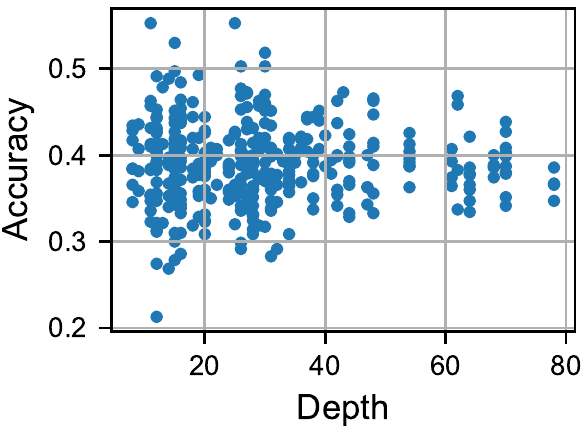}
\caption{Accuracy vs. depth for $p=8$~\pcs.}
\label{glass_pca_8_depth_acc}
  \end{subfigure}
  \begin{subfigure}[t]{150pt}
\includegraphics[width=1\textwidth]{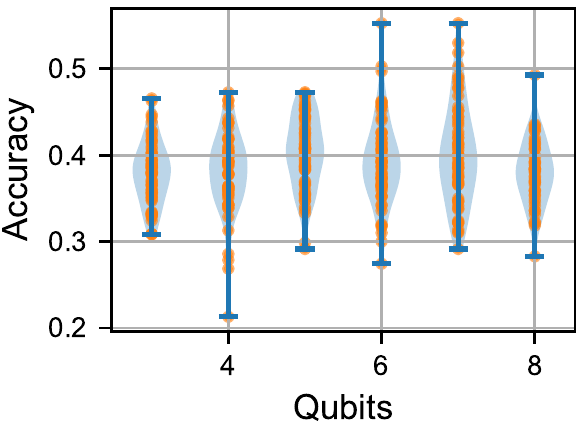}
\caption{Accuracy vs. No. qubits for $p=8$~\pcs.}
\label{glass_pca_8_qubits}
  \end{subfigure}
  \begin{subfigure}[t]{150pt}
\includegraphics[width=1\textwidth]{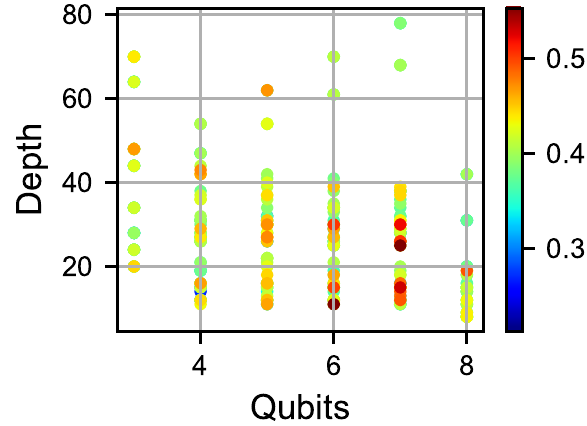}
\caption{Depth vs. No. qubits colour-coded by Accuracy for $p=8$~\pcs.}
\label{glass_pca_8_depth_bits}
  \end{subfigure}
  \hfill %%
  \caption{\label{fig:glass_pca_8}Relationship between accuracy, number of qubits, and depth on the first $8$ principal components (PCs) of the glass dataset. Each point represents the mean of 10 training runs without noise.}
  \vspace{-.3cm}
\end{figure*}

\begin{figure*}[htbp]
\centering
  \begin{subfigure}[t]{150pt}
    \includegraphics[width=1\textwidth]{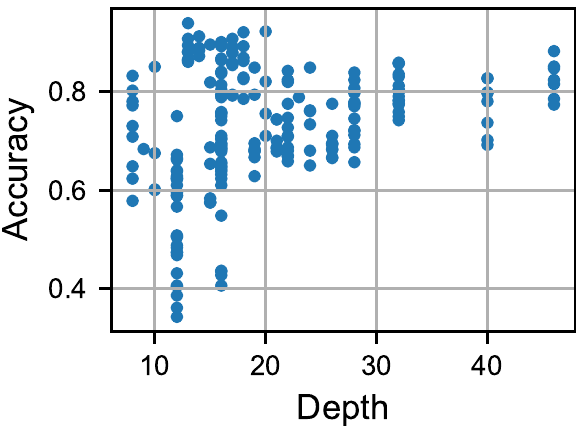}
\caption{Accuracy vs. depth.}
\label{iris_depth_acc}
  \end{subfigure}
  \begin{subfigure}[t]{150pt}
\includegraphics[width=1\textwidth]{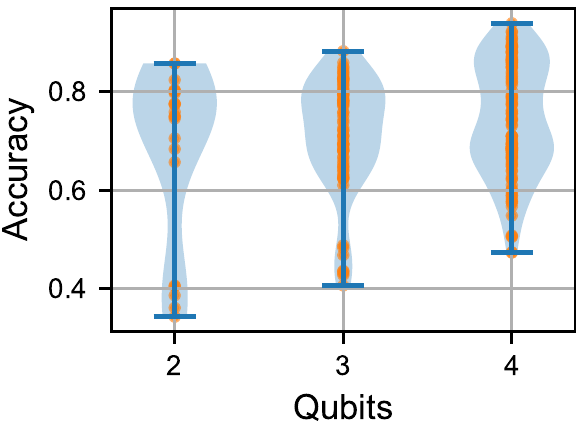}
\caption{Accuracy vs. No. qubits.}
\label{iris_qubits}
  \end{subfigure}
  \begin{subfigure}[t]{150pt}
\includegraphics[width=1\textwidth]{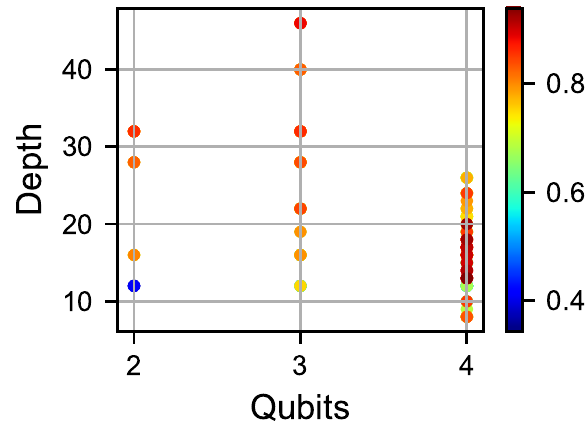}
\caption{Depth vs. No. qubits colour-coded by Accuracy.}
\label{iris_depth_bits}
  \end{subfigure}
  \hfill %%
  \caption{\label{fig:iris}Relationship between accuracy, number of qubits, and depth of the iris dataset. Each point represents the mean of 10 training runs without noise.}
  \vspace{-.3cm}
\end{figure*}

\begin{figure*}[hp]
\centering
\includegraphics{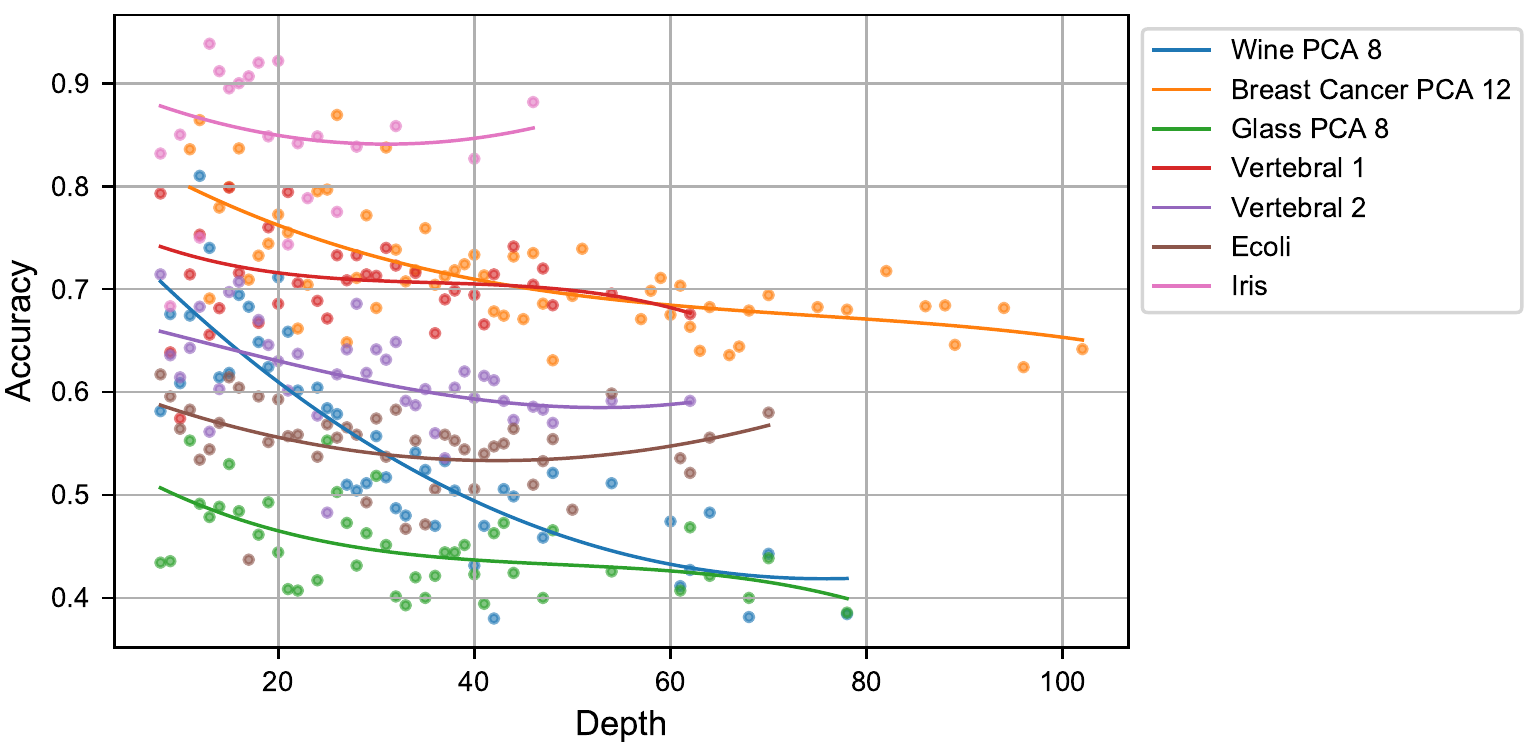}
\caption{Accuracy vs. circuit depth. Each point represents the mean accuracy of the best circuit at a given depth from training runs without noise. Each line represents the trend of the data by fitting a cubic spline.}
\label{result_comparison}
\vspace{-.5cm}
\end{figure*}

\begin{figure*}[hp]
\centering
\includegraphics{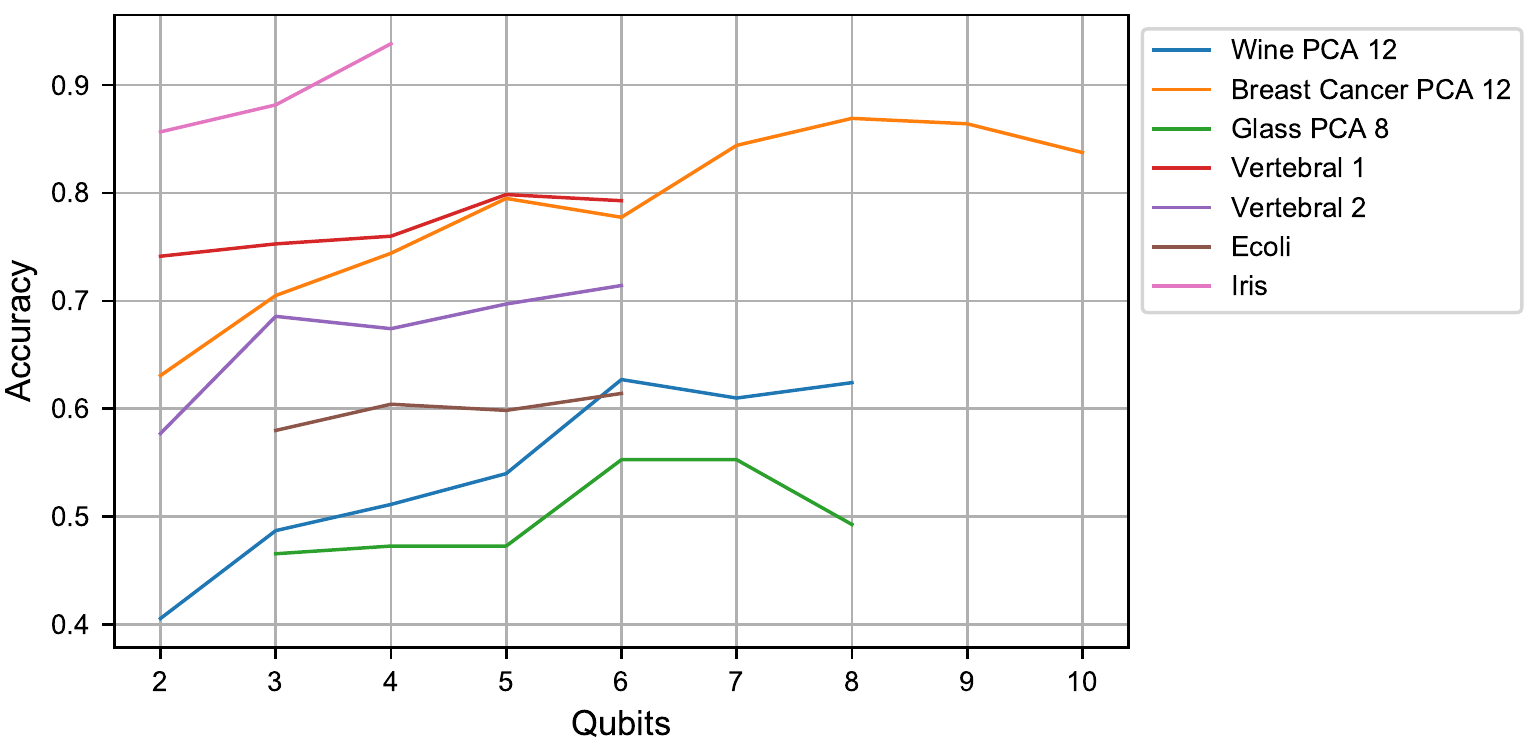}
\caption{Accuracy vs. number of qubits. Each line represents the mean accuracy of the best circuit with $n$ qubits from training runs without noise.}
\label{acc_vs_bits}
\vspace{-.5cm}
\end{figure*}

\killsubsectionspace
The results of the experiments were analyzed considering the number of qubits used and the depth of each circuit with respect to model accuracy. We first highlight specific findings using different datasets (Figures \ref{fig:wine}--\ref{fig:iris}) in more detail and then make some general comments for all datasets (\autoref{result_comparison} and \ref{acc_vs_bits}). In Figures \ref{fig:wine}--\ref{acc_vs_bits}, each data point represents the mean performance (accuracy) of 10 independently trained circuits on 10 independent unseen test sets. The differences are due to randomness in the initial values for the parameters. Using the mean reduces the influence of the choice of these parameters on the results and makes the results thereby more dependent on the topology of the circuit.

The experimental results of training a variational quantum classifier on the first 12~\pcs of the Wine dataset with 307 unique circuits is shown in \autoref{wine_pca_12_depth_acc}. The data shows that the accuracy upper bound tends to decrease as circuit depth increases. At the same time, the overall variance in the results decreases. The influence of the number of qubits on the accuracy can be seen in \autoref{wine_pca_12_qubits}. We observe that accuracy increases with the number of qubits until the accuracy plateaus. The plateaus in \autoref{wine_pca_12_qubits} are achieved when the model consists of as few layers as possible as can be seen in \autoref{wine_pca_12_depth_bits}. However, circuits with fewer qubits must be deeper as each feature must be applied at least once (see~\autoref{ssec:circuit_depth}).

\autoref{fig:breast_cancer} depicts the same plots for the first 8~\pcs of the Breast Cancer Wisconsin dataset. Overall the highest accuracy can be found for low depths in \autoref{breast_cancer_pca_12_depth_acc} but the lowest accuracy as well. The plot shows that the overall variance in the accuracy is highest for low depths whereas higher depth circuits show a lower variance in the achieved accuracy, but also a lower accuracy. The highest accuracy is achieved with 8 qubits as can be seen in \autoref{breast_cancer_pca_12_qubits} but the performance drops for 9 and 10 qubits. \autoref{breast_cancer_pca_12_depth_bits} shows that the best circuits are found for low depth and a higher number of qubits.

The results on the first 8~\pcs of the glass dataset show a similar behaviour as for the breast cancer dataset (\autoref{fig:glass_pca_8}). The accuracy is the highest for a small depth and a high number of qubits, see \autoref{glass_pca_8_depth_bits}, but the effect is not as pronounced as for the other two datasets. The highest accuracy starts dropping when increasing the number of qubits beyond 7, see \autoref{glass_pca_8_qubits}.

\autoref{fig:iris} shows the experimental results on the iris dataset. The best accuracy is still achieved for a low circuit depth, \autoref{iris_depth_acc}, but the accuracy for a higher depth is only not a lot lower while the variance is higher for lower depths overall. The higher variance is especially pronounced in the lower accuracy results. While higher depth circuits all tend to achieve a smaller variance in the accuracy. A higher number of qubits is still able to achieve the best results, \autoref{iris_qubits} and \autoref{iris_depth_bits} show that the best circuits still have low depth and high number of qubits even though the effect is not as pronounced. Low depth and a low number of qubits give the worst results.

Overall across the datasets illustrated (note we also observe similar patterns in the remaining datasets noted in \autoref{DatasetTable}), the lower the depth circuits tend to show a higher variance in the accuracy while at the same time achieving the overall best accuracy. At the same time, the best accuracy is achieved when a higher number of qubits is used. 
A possible explanation here might be that smaller circuits often have more qubits which has more combinations on how the $CZ$-gates can be inserted between the parameterized gates. This increased variance at lower depths and higher numbers of qubits means that although the best circuits are often found in these portions of the design space, that there are no guarantees that a shallow and wide circuit will perform well, but rather it indicates that this region of the design space is more likely to have a ``good'' circuit in terms of classification accuracy.


Generalising across the studied datasets, \autoref{result_comparison} shows the accuracy of the best circuits at given depths. The overall trend in the results for each dataset is approximated by a cubic spline. We see that average accuracy decreases the most quickly for the Wine dataset. A decrease in the upper bound of the accuracy as the depth is increased can also be found for the Vertebral 1, Glass and Wisconsin Breast Cancer dataset. In contrast, the Iris and Ecoli datasets start to increase again after an initial drop in performance. In our current design setup it is not possible to increase the number of layers further without repeating the features more than twice. Both were not able to achieve the best accuracy with deeper circuits, however. The Vertebral 2 dataset behaves similarly. Overall the best accuracy is observed across all datasets with shallower circuits.

We can also observe that there is a tendency to perform better with a higher number of qubits across all datasets (see \autoref{acc_vs_bits}). Two notable exceptions are  the Glass and Wisconsin Breast Cancer datasets. In these cases, the performance starts to drop as the number of qubits is increased further. This means that increasing the number of qubits can be beneficial but there is no guarantee that the best accuracy be achieved when the number of qubits is equal to the number features. All results are with and without repeated use of features, therefore we could show no benefit to a repetitions of features in the circuits as in our current setup this would require an increase in the number of layers. 

\subsection{Challenges of High(er) Dimensional Data}
\killsubsectionspace
While we evaluated variational quantum classifiers on a range of datasets containing different number of features, we also evaluated the impact of the number of features on classifier accuracy. We use PCA as a preprocessing step and then use a varying number of features (\pcs) in the training. An alternative could be to take increasingly large subsets of features. However, as not all features are equally informative, this might mean that a very (un)informative feature is added, which would not be easy to account for and potentially result in a method bias. Using PCA as prepossessing is more suitable as it gives us a relative ranking of the derived features (\pcs) with regard on how informative they are. This way a model that receives the first $n$ \pcs receives the most informative features and adding additional \pcs should add a relatively smaller amount of information. A circuit using $n+1$ \pcs should in principle be at least as good as a model using $n$ \pcs, as only information is added.

\begin{figure*}[ht]
\centering
  \begin{subfigure}[b]{230pt}
    \includegraphics[width=\textwidth]{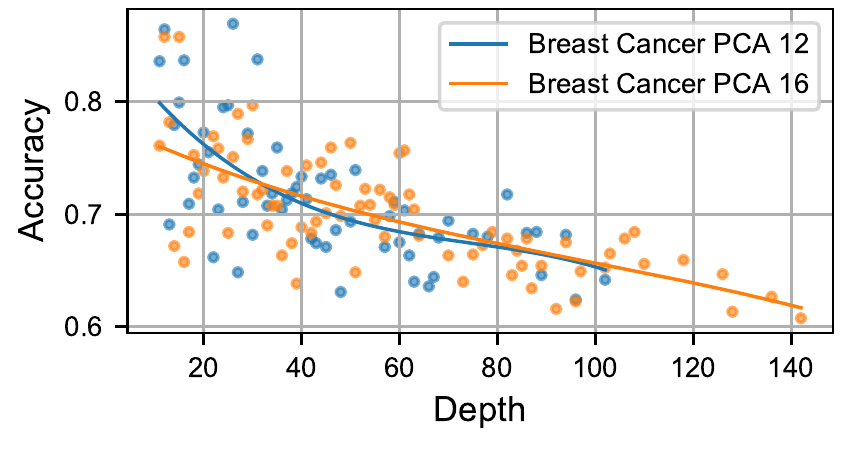}
\caption{Breast Cancer dataset with $12$ \pcs and $16$ \pcs.}
\label{cancer_pca_12_16}
  \end{subfigure}
 % \hfill %%
  \begin{subfigure}[b]{230pt}
   \includegraphics[width=\textwidth]{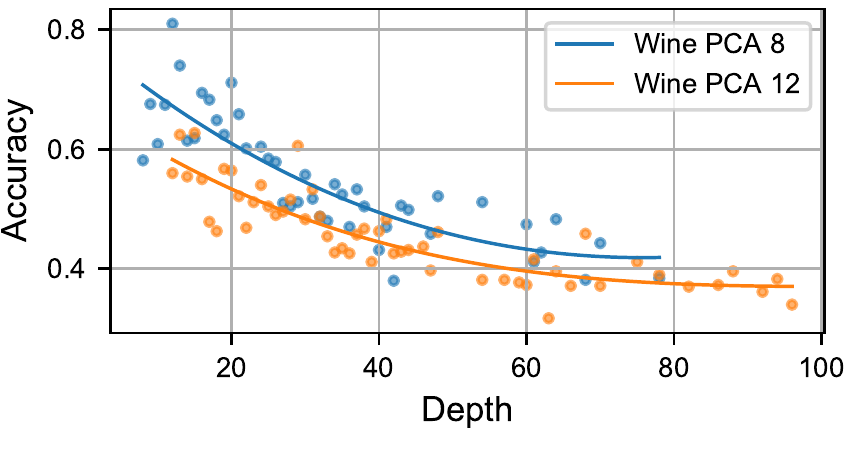}
\caption{Wine dataset with $8$ \pcs and $12$ \pcs.}
\label{wine_pca_8_12}
  \end{subfigure}
  \hfill %%
  \caption{\label{fig:num_features}Influence of number of \pcs. 
  %Accuracy on the y-Axis and Circuit depth on the x-Axis. 
  Each point represents the mean accuracy of the best circuit at one depth on the first $n$ \pcs from training runs without noise. Each line represents the trend of the data by fitting a cubic spline to it.}
  \vspace{-.15cm}
\end{figure*}

However, exploring the Wisconsin Breast Cancer dataset (see \autoref{cancer_pca_12_16}) we see that the best circuits trained using 12 \pcs achieve a higher accuracy than the circuits that trained using 16 \pcs at shallow(er) depths, and as depth increases, models converge to a similar accuracy. It is important to note that both 12 and 16 \pcs allow the same minimal depth. 
\autoref{wine_pca_8_12} shows the same comparison for the wine dataset but with 8 and 12 \pcs. Here, a fast decrease in accuracy can be observed as depth increases but the circuits trained on 8 \pcs generally perform better. The best accuracy for circuits trained on 8 \pcs was 81\% but only 63\% with 12 \pcs . The Wine dataset actually proved quite challenging (suggesting it be used as a benchmark in other works focusing on \vqml or other methods of training quantum machine learning methods) even though classical models can achieve 100\% accuracy \cite{wine_acc}. 

% \begin{figure}[h]
% \centering
% \includegraphics[width=.45\textwidth]{figures/cancer_pca_12_16.png}
% \caption{Accuracy on the y-Axis and Circuit depth on the x-Axis. Each point represents the mean accuracy of the best circuit at one depth on the first 12 and 16 PCA components of the Wisonsin breast cancer dataset. Each line represents the trend of the data by fitting a cubic spline to it.}
% \label{cancer_pca_12_16}
% \end{figure}

% \begin{figure}[h]
% \centering
% \includegraphics[width=.45\textwidth]{figures/wine_pca_8_12.png}
% \caption{Accuracy on the y-Axis and Circuit depth on the x-Axis. Each point represents the mean accuracy of the best circuit at one depth on the first 8 and 12 PCA components of the Wine dataset. Each line represents the trend of the data by fitting a cubic spline to it.}
% \label{wine_pca_8_12}
% \end{figure}
\subsection{Summary}
\killsubsectionspace
 The results suggest that the best performance may be achieved when circuits are shallow, i.e. shallow circuits seem to have a higher upper bound of achievable accuracy. To minimise depth, our analysis suggests that practitioners should design circuits %such that the product of qubits and feature blocks which is bigger or equal to the number of features should minimize the depth. This means that the circuit should
with at least as many qubits as there are features (or \pcs) in the dataset; %or at least as close as possible based 
obviously cognisant of hardware constraints. 
%This was shown, experimentally, to achieve best results.

Our results suggest that circuits tend to perform worse as the number of features increases, even if we scale the number of qubits accordingly. This is a problem that needs to be addressed in order to scale this approach for the training of \vqml classifiers for datasets containing more features. In this paper, we explore mainly ``small'' and ``easy'' datasets, yet our results indicate that with today's hardware availability it is not (yet) feasible to set the number of qubits equal to the total number of features. However, it is also worth noting that so far we have explored only error free computations. As such, these results are a best case scenario, as many modern day quantum computers are not error free (NISQ), but have aspects of ``noise'' in their operations and measurement. 

\section{Noisy Circuit Performance}
\label{sec:noise_results}
\killsectionspace
In this section we present the results of training circuits when the simulated quantum computer exhibits various noise characteristics in the $CZ$ and measurement operations, resulting in errors in those computations. This is done to evaluate the overall robustness of our observations in \autoref{section:Noise_free_results} to provide a commentary on the design of \vqml models in the presence of errors on NISQ QCs. This is important since the errors in the computations, with a bias towards relaxed states, result in an error in the computed loss function, which is used to train the model (circuit). This might make the overall optimization of the circuit more difficult. It is therefore important to analyze how (and if) the training is able to cope with these errors. 
We find that the measurement noise is independent of the circuit depth whereas errors in the $CZ$-gate affect the performance of deeper circuits more as they tend to contain more $CZ$-gates.

\begin{figure*}[ht]
\centering
  \begin{subfigure}[b]{230pt}
    \includegraphics[width=\textwidth]{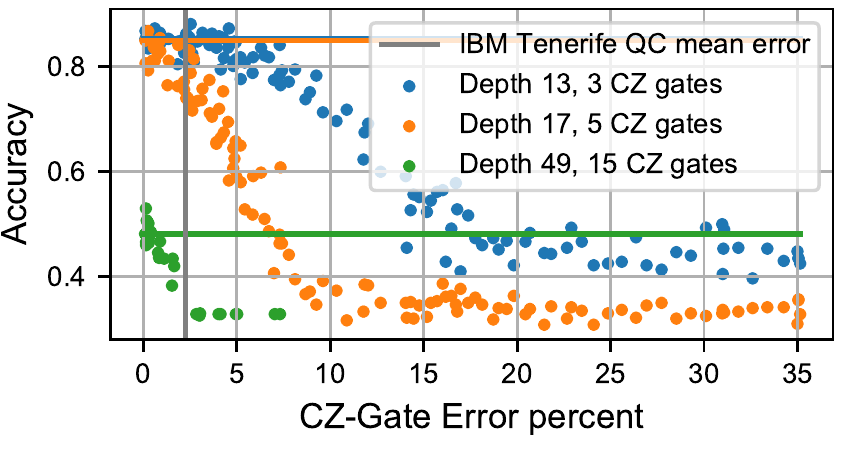}
\caption{Accuracy vs $CZ$-error percentage.}
\label{$CZ$_noise}
  \end{subfigure}
%  \hfill %%
  \begin{subfigure}[b]{230pt}
    \includegraphics[width=1\textwidth]{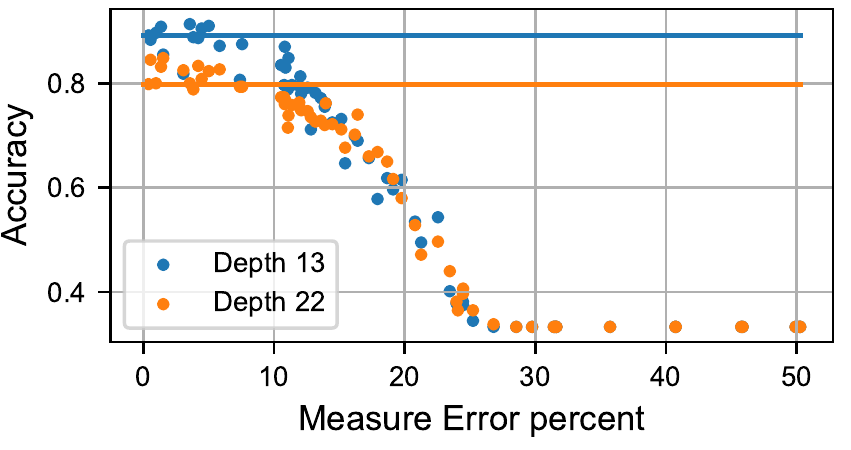}
\caption{Accuracy vs \emph{Measurement} error percentage.}
\label{measure_noise}
  \end{subfigure}
  \hfill %%
  \caption{\label{fig:accuracy}Accuracy vs. $CZ$-(left) and \emph{Measurement} error (right) error for circuits of varying depth trained on the Iris dataset. There horizontal lines represent the mean accuracy of models exectued without noise and the vertical grey line the error that the IBM Tenerife QC applies in $CZ$-gates. \cite{ibm_error}}
  \vspace{-.25cm}
\end{figure*}

To estimate the influence of noise on the results $n=69$ circuits were evaluated on the Iris data set, with errors in the $CZ$-gate computation ranging from 0\% to 35\% and the measurement error from 0\% to 50\%. \autoref{$CZ$_noise} shows a comparison of the accuracy of circuits with a depth of 13, 17 and 49 when thermal relaxation is applied to the $CZ$-gate. Note that 0\% error would be represent a circuit trained using the same approach from the previous section. In this section, we take a number of fixed circuit designs and train them under a series of error rates, i.e. the circuit designs are fixed, but the error rates change. Thus illustrating the impact on \vqml model performance of errors due to thermal relaxation and noise in measurements. While, we graphically illustrate only a few examples, the general trend of the results hold across the circuits studied. 

\subsection{Error in $CZ$-gate computations}
\killsubsectionspace

% \begin{figure}[h]
% \centering
% \includegraphics[width=.45\textwidth]{figures/$CZ$_noise.png}
% \caption{Accuracy on the y-Axis and Circuit depth on the x-Axis. Each point represents the mean accuracy of the best circuit at one depth on the first 12 and 16 PCA components of the Wisonsin breast cancer dataset. Each line represents the trend of the data by fitting a cubic spline to it.}
% \label{$CZ$_noise}
% \end{figure}

%The deepest circuit has the lowest initial accuracy and it sharply drops with increasing error in the $CZ$-gate. 

The accuracy of the circuit with a depth of 17 drops when increasing the $CZ$-gate error from 0\% to 10\%. At 10\% the accuracy of the circuit starts to plateau at an accuracy of 0.35 (note that for this data set 33\% accuracy would be equivalent to just picking a class at random). The performance of the circuit with a depth of 13 starts dropping at roughly 5\% and reaches an accuracy of 0.45 at a gate error of 35\%. The biggest difference between the depth 13 and 17 circuits is the number of $CZ$-gates as both have the same number of rotations. This means that decrease in accuracy is due to the number of $CZ$-gates on the longest path which increases as circuit depth increases. The grey vertical line shows the percentage of errors introduced in the $CZ$-gates on the the IBM Tenerife QC; giving an indication of what modern day hardware exhibits. At this error the two circuits with a depth of 13 and 17 show good performance. The error of the QC was displayed as example since it exhibits the highest error of all QCs reported by IBM.

The circuit with a depth of 49 does already start-off worse when no error is present as can be seen from the green horizontal line. The accuracy directly starts dropping as the error in the $CZ$-gate is increased. It reaches its minimal value at 2.7\% $CZ$-gate error. At the expected error rate on IBM Tenerife QC it would not learn to classify the dataset as the accuracy is not better than random chance. The reason for this is probably the high number of $CZ$-gates, used on the longest path through the circuit, as the error in the computations accumulates.

%\subsection{Error in \emph{Measurement}-gate computations}
\subsection{Measurement error}
\killsubsectionspace

The effects of measurement noise from 0\% to 50\% can be seen in \autoref{measure_noise}. Initially the deeper circuit starts with a lower accuracy but both circuits 
%start to drop at the same error and reach a similar accuracy at the same error percentage after the initial drop. As the initial difference between the circuits is not due to noise. 
behave similarly (in terms of accuracy decay) as the measurement error increases. Whilst we show only two circuits here, in general, we observed similar trends for other circuits. Thus, it appears that decays in accuracy due to measurement noise is independent of circuit depth; more experimentation would be needed to confirm this. %Both circuits start dropping sharply, in accuracy, when noise exceeds $\approx 10\%$. 

\subsection{Summary}
\killsubsectionspace

Low depth circuits are more robust against the influence of errors in the $CZ$-gate quantum computations, i.e. 
%through simulation, we see that machine learning models developed with 
shallow circuits appear more robust against thermal relaxation errors. The influence of measurement noise is independent of the depth of the circuit.
At present, our results suggest that a quantum computer with a high(er) number of low(er) quality qubits could be advantageous for \vqml classifiers. %In order to draw such conclusions further studies on the resilience of variational quantum classifier towards noise is necessary in more depth incorporating more diverse noise model.

 %Additionally, the variance in 

%% file: tex_files/Conclusion.tex
\section{Limitations and Discussion}
\label{sec:limitations} 
\killsectionspace
The observations made above are not complete, and also have some caveats that should be explored in future work to establish comprehensive set of guidelines beyond those we introduce in the next section. Overall the variance in performance is often high in ``good'' regions of the design space, and thus design guidelines are needed to reduce the variance of the results further and achieve more consistent results comparable to classical approaches. This is especially important since low depth circuits exhibited a larger variance eventually due to the possible $CZ$-Gate configurations. It must also be explored whether a further increase in the number of qubits would improve the results further. 

It is important to stress that circuits should not be arbitrarily wide, as in order to leverage larger numbers of qubits, more entanglement operations (like the $CZ$-gate) are needed, and this naturally increases the minimum depth of the circuit and the variance in $CZ$-gate configurations. Our design space and constraints on its exploration limit the maximum number of qubits leveraged. Relaxing these constraints, e.g. by using features multiple times, would allow for wider circuits, however, as noted above this could also increase circuit depth and potentially variance. Finding inflection points in this trade-off would be useful. There is also the question how influential the barren plateau problem \cite{barrenplat} is on our findings, suggesting further study of various circuit designs.% A study exploring the influence of it given various circuit designs is therefore necessary. %Thus when interpreting the observations above and any guidelines resulting from future work, care and common sense is needed.

As the number of noise models applied in this work is quite small, future research should explore a larger variety of noise models and sources of noise. This would help find noise resilient circuits early on that can then be further evaluated on real devices which is a crucial step for further development. Doing this would require taking into account the influence of the hardware on which it is executed and the transpilation algorithm applied \cite{JohnMcAllister}. %Studies evaluating the influence of transpilation on the performance on real devices are therefore necessary and the design of the circuits . 
%\item Our results and correspondingly are derived using simulation. In general, next steps would also be to verify our findings on available quantum hardware.
The overall design of the \vqml circuits should best be done taking both factors available hardware and transpilation into account. 

Finally, based on our results it would be tempting to try and derive observations on the basis of class cardinality and number of features a dataset has, i.e. are classification problems with more classes / features ``easier'' or ``harder''? However, caution is needed here, our results despite exploring a large number of circuits, still only represent 5 datasets, and thus we would not be able to reliably answer such a question, instead, an experimental design similar to \cite{100datasets} would be needed (which explores larger numbers of datasets and models and devises an evaluation methodology for such settings).

\section{Conclusion}
\label{section:conclusion}
\killsectionspace
This paper aims to establish some initial design guidelines for \vqml classifiers based on experiments with a large number ($n=6500$) of circuits applied to five common datasets in the machine learning literature. We observe that shallow circuits are more robust to $CZ$-gate errors and generally have higher accuracy potential (both with and without errors). It is also easier to have shallow(er) circuits if they are wide (up to the number of features used), but that the number of features used should be small. Where more features are present, the number of features used can be constrained by using dimensionality reduction techniques like PCA. We also observed that Measurement error is only a significant factor if ``large'', i.e. $>10\%$. Thus, our results suggest four design guidelines for variational quantum machine learning (\vqml):% circuits: 

\begin{enumerate}
    \item Circuits should be as shallow as possible, AND
    \item Circuits should be as wide as possible, AND
    \item The number of features should be kept small, BUT
    \item Small amounts of noise can be tolerated.
\end{enumerate}

%The accuracy drop due to noise in the measurement is independent of the depth and the circuits are fairly robust towards as it is not dropping much below an error of 10\%. 

As a key takeaway, we observe that quantum computers with a high(er) number of low quality qubits could be advantageous for \vqml classifiers but more research is needed to assess whether more substantial numbers of low(er) quality qubits will further improve performance.
%It is also not clear yet whether variational quantum classifiers offer any quantum advantage over classical approaches.

% Structure:
% \begin{itemize}
%     \item Main findings: positives of shallow, errors not a problem up to a certain extent etc.
%     \item Main open challenges (see list below)
%     \item Future work
% \end{itemize}
% Having more bits for a given tasks tends to improve the overall results while not increase the influence of errors in the computation.

% Shallow circuits are more robust towards errors in the computations.

% Shallow circuits seem to perform better than deeper ones.

% The optimizer has an non-negligible impact on the result but none seem to work well on very deep circuits.

% - To solve a specific problem a circuit with many bits (as much as the number of features) but low depth might be beneficial.
% - Hardware manufacturers might find an audience for noise Quantum Computer with high bit numbers.

% The open challenges for this manner of addressing machine learning with QC:
% \begin{itemize}
%     \item class cardinality and bit string width
%     \item number of features vs. circuit
%     \item handling high(er) variance
%     \item choice of optimiser
% \end{itemize}